\documentclass[
	pra,
	twocolumn,
	superscriptaddress,
	longbibliography
]{revtex4-2}

\usepackage[T1]{fontenc}
\usepackage[utf8]{inputenc}
\usepackage[dvipsnames,table]{xcolor}
\usepackage{graphicx}
\usepackage{latexsym}
\usepackage{amsmath}
\usepackage{graphics}
\usepackage{amssymb}
\usepackage{layout}
\usepackage{verbatim}
\usepackage{amsfonts,epsfig,dsfont}
\usepackage{braket}
\usepackage{soul}
\usepackage{hyperref}
\usepackage{bbm,bm}

\hypersetup{
	colorlinks,
	linkcolor={blue!90!black},
	citecolor={blue!90!black},
	urlcolor={blue!90!black}
}

\newcommand{\Tr}{\textrm{Tr}}

\newcommand{\JS}[1]{\textcolor{magenta}{#1}}

\begin{document}

\title{
	Qubit-environment entanglement in pure dephasing dynamics for mixed initial states
}

\author{Ma\l{}gorzata Strza\l{}ka}
	\affiliation{FZU - Institute of Physics of the Czech Academy of Sciences, Na Slovance 2,  182 00 Prague, Czech Republic}
\affiliation{Faculty  of  Mathematics  and  Physics, Charles University, Ke Karlovu 5,
121 16 Prague, Czech Republic}
\author{Nico Peters}
	\affiliation{Theoretical Quantum Science, Institute for Photonic Quantum Systems (PhoQS), Paderborn University, Warburger Stra\ss{}e 100, 33098 Paderborn, Germany}

\author{Laura Ares}
	\affiliation{Theoretical Quantum Science, Institute for Photonic Quantum Systems (PhoQS), Paderborn University, Warburger Stra\ss{}e 100, 33098 Paderborn, Germany}

\author{Jan Sperling}
	\affiliation{Theoretical Quantum Science, Institute for Photonic Quantum Systems (PhoQS), Paderborn University, Warburger Stra\ss{}e 100, 33098 Paderborn, Germany}

\author{Katarzyna Roszak}
	\affiliation{FZU - Institute of Physics of the Czech Academy of Sciences, Na Slovance 2,  182 00 Prague, Czech Republic}

\date{\today}

\begin{abstract}
	Decoherence of qubits is one main limitation in quantum information processing, where the qubits themselves can be mixed because of decoherence and additional decoherence may generated through interactions with a bath.
	Here, we present a complete study of qubit-environment entanglement in the case of interactions that lead to pure dephasing of the qubit by allowing the initial state of the qubit to be mixed, exceeding previous restrictions.
	We find that the criteria for separability that hold for pure initial states of the qubit are reduced to witness status.
	Yet, the evolution of the witnesses reflects the array of features that entanglement can exhibit in the mixed case but are forbidden in the pure-qubit case.
	Most notably, this includes sudden birth of entanglement as the initial phase of the evolution is typically separable.
	Moreover, there is a stark numerical advantage to the study of entanglement for large environments when using the witness rather than the traditional measures available.
	We include the study of entanglement generated between a spin qubit and a sparse spin environment to illustrate the peculiarities of the evolution of entanglement in this case, as well as to emphasize the level of reliability of the entanglement witness.
\end{abstract}

\maketitle

\section{Introduction}
\label{sec1}

	The general description of qubit-environment entanglement (QEE), assuming an arbitrary size of the environment, is numerically demanding for mixed states and either requires minimization over a large number of parameters or the diagonalization of large matrices \cite{Plenio_QIC07}.
	Nevertheless, if the interaction is limited to Hamiltonians that can only lead to pure dephasing of the qubit, then the distinction between an entangled and separable qubit-environment state can be made analytically as long as the initial state of the qubit is pure \cite{roszak15a}, while the state of the environment can be arbitrary.

	The existence of easily accessible criteria for QEE rendered advancements possible that could not be facilitated by numerical studies alone.
	Primarily, for pure dephasing, there is a physical signature of the environment that distinguishes decoherence as the result of entanglement from decoherence as the result of the buildup of classical correlations \cite{roszak15a}.
	This signature can be measured by operations and measurements on the qubit alone, using back-action \cite{roszak19a,rzepkowski20i}, and the presence of QEE can have nontrivial effects on the operation of quantum algorithms \cite{roszak21,harlender22,roszak23, roszak23,strzalka_arxiv}.
	Taking into account the fact that many platforms currently considered for quantum computation are based on solid-state qubits, and that the predominant decoherence mechanisms lead to pure dephasing for such qubits \cite{Yang_RPP17,Malinowski_PRL17,deLange_Science10,knee16,salamon17_i,bravyi22,zhao22,anferov24,fung24,beukers25}, the effect of entangling noise on the operation of quantum algorithms is highly relevant.

	In this paper, we close the gap to a complete description of QEE for evolutions that lead to pure dephasing of the qubit by relaxing the main assumption made in Ref. \cite{roszak15a};
	namely, the initial qubit does not have to be in a pure state in this work.
	Unsurprisingly, we find that an interaction that cannot entangle a pure initial qubit state with its environment also does not result in entanglement buildup from mixed qubit states.
	For entangling evolutions, however, the situation becomes much richer, and we find that the set of if-and-only-if conditions for qubit-environment separability reduces to an entanglement witness for initially mixed qubit states.
	This means that when the conditions certify entanglement, the state is indeed entangled, but separability is more ambiguous.

	Remarkably, features in the evolution of entanglement that are impossible for pure initial qubit states become possible and even common for initially mixed states.
	The most significant additional effect is the sudden birth of entanglement \cite{ficek08,mazzola09}.
	If an evolution is entangling, QEE for pure states starts forming immediately, and only isolated points in time of separability over the evolution occur \cite{roszak17i}.
	Entanglement for initially mixed qubit states, by contrast, typically involves separability for an initial period, followed by the buildup of entanglement.
	In oscillatory systems, a sudden death is often followed by a sudden rebirth of entanglement.

	To demonstrate the connection to contemporary experimental systems, we study in detail an example consisting of a nitrogen-vacancy (NV) center spin qubit interacting with a sparse environment of nuclear spins interacting via the hyperfine interaction \cite{ jelezko06, gali08, zhao12_d, doherty13, kwiatkowski18, wood18i, strzalka20, strzalka21,finsterhoelzl25}.
	This serves to highlight the possible discrepancies between the actual evolution of QEE and the witnessed presence of such entanglement.
	While the general qualitative behavior present and detected entanglement is typically similar when using the witness for strong entanglement, it is particularly unreliable when entanglement is weak.
	Nevertheless, in many situations, it remains a practical tool for the study of QEE since, contrary to measures of entanglement, it does not require the tomographic reconstruction and diagonalization of large matrices.

	The remainder of the paper is organized as follows.
	Section \ref{sec2} reviews the methodology used to describe the full system-environment density matrix under the influence of a Hamiltonian that leads to pure dephasing of the qubit.
	Section \ref{sec3} recapitulates the criteria for QEE for pure initial states.
	Section \ref{sec4} contains a study of the generation of entanglement for pure dephasing evolutions for mixed initial states.
	Proof-of-concept evolutions of entanglement for a spin qubit interacting with a few spin environment via the hyperfine interaction are discussed in Sec. \ref{sec5}.
	Section \ref{sec6} eventually concludes the paper.

\section{Pure dephasing evolutions}
\label{sec2}

	We consider a single qubit in the presence of an environment in the situation where their interaction can only lead to pure dephasing of the qubit.
	Such interactions are often the dominant decoherence mechanisms for leading solid-state qubits, at least in some ranges of external parameters \cite{Yang_RPP17,Malinowski_PRL17,deLange_Science10,knee16,liao10,salamon17_i,bravyi22,zhao22,anferov24,fung24,beukers25}.
	The characteristic feature of such interactions is that the full qubit-environment Hamiltonian has separable eigenstates and, as such, it is easier to consider it in general terms \cite{roszak15a,roszak18,chen18,chen19}.

	Any Hamiltonian that leads to pure dephasing of the qubit can be written in the general form
	\begin{equation}
		\label{ham}
		\hat{H} = \sum_{i\in\{0,1\}} \varepsilon_i |i\rangle \langle i | + \hat{H}_E + \sum_{i\in\{0,1\}} |i\rangle \langle i | \otimes \hat{V}_i.
	\end{equation}
	Therein, the first term of the Hamiltonian characterizes the qubit, the second term is the free Hamiltonian of the environment, and the last term specifies the qubit-environment interaction.
	This form guarantees the commutation between the free qubit Hamiltonian and the interaction term, which excludes processes involving energy exchange between the qubit and the environment.
	Thus, there exists a qubit basis that diagonalizes the whole Hamiltonian, and, in this way, the interaction specifies qubit pointer states \cite{Zurek_PRD82,zurek03i}.
	Here, these are denoted as $|i\rangle$, with $i\in\{0,1\}$, in Eq. \eqref{ham}.
	There are neither restrictions on the environmental part $\hat{H}_E$ nor the conditional operators $\hat{V}_i$, which describe the effect that the qubit state has on the environment.

	For the aforementioned class of Hamiltonians, the qubit-environment evolution operator decomposes as follows \cite{roszak18}:
	\begin{equation}
		\label{ewolucja}
		\hat{U}(t) = \sum_{i\in\{0,1\}} |i\rangle \langle i | \otimes\hat{w}_i(t),
	\end{equation}
	where the evolution of the environment when the qubit is in state $|i\rangle$ is described by the conditional evolution operators $\hat{w}_i(t)$, given by the unitary maps
	\begin{equation}
		\label{w}
		\hat{w}_i(t) =
		e^{-i\varepsilon_i/\hbar}
		\exp\left[-\frac{i }{\hbar}\left(
			\hat{H}_E + \hat{V}_i
		\right)t\right].
	\end{equation}

	The above formula enables us to formally find the joint state at a given time $t$ for any initial qubit-environment state.
	As a reasonable assumption, we restrict our considerations to uncorrelated initial states,
	\begin{equation}
		\label{ini}
		\hat{\sigma}(0)=\hat{\rho}(0)\otimes\hat{R}(0),
	\end{equation}
	where the initial states of both the qubit, $\hat{\rho}(0)$, and the environment, $\hat{R}(0)$, are arbitrary.
	For the qubit, we consider the decomposition
	\begin{equation}
		\label{iniq}
		\hat{\rho}(0)=\left(
		\begin{array}{cc}
			|a|^2&ab^*\varepsilon
			\\
			a^*b\varepsilon&|b|^2
		\end{array}
		\right),
	\end{equation}
	in which the decoherence factor $\varepsilon\in[0,1]$ can be assumed real without loss of generality, implying that all phase information is contained in the complex coefficients $a$ and $b$.
	The joint qubit-environment (QE) state, after propagation for a time $t$, is then given in a block form (i.e., operator-valued matrix entries) as
	\begin{equation}
		\label{sigma}
		\hat{\sigma}(t)=\left(
		\begin{array}{cc}
			|a|^2\hat{R}_{00}(t)&ab^*\varepsilon\hat{R}_{01}(t)
			\\
			a^*b\varepsilon\hat{R}_{10}(t)&|b|^2\hat{R}_{11}(t)
		\end{array}
		\right),
	\end{equation}
	utilizing the shorthand notation
	\begin{equation}
		\label{rij}
		\hat{R}_{ij}(t)=\hat{w}_i(t)\hat{R}(0)\hat{w}_j^{\dagger}(t).
	\end{equation}
	Here, the state is written in matrix form only in terms of qubit pointer states by convention.
	In the following, we will always assume that $a\neq 0$ and $b\neq0$, otherwise dephasing does not occur and entanglement is not formed.

	It is often convenient to write the QE state in terms of $\hat{R}_{00}(t)$, i.e., the state of the environment conditioned on the pointer state $|0\rangle$,
	\begin{equation}
		\label{sigma2}
		\hat{\sigma}(t)=\left(
		\begin{array}{cc}
			|a|^2\hat{R}_{00}(t)&ab^*\varepsilon\hat{R}_{00}(t)\hat{w}^{\dagger}(t)
			\\
			a^*b\varepsilon\hat{w}(t)\hat{R}_{00}(t)&|b|^2\hat{w}(t)\hat{R}_{00}(t)\hat{w}^{\dagger}(t)
		\end{array}
		\right).
	\end{equation}
	This simplifies the analysis because it solely requires one environmental operator,
	\begin{equation}
		\label{w2}
		\hat{w}(t)=\hat{w}_1(t)\hat{w}_0^{\dagger}(t).
	\end{equation}

\section{QEE for pure initial qubit states}
\label{sec3}

	The problem of QEE quantification of any state of the form \eqref{sigma} as long as the initial state is pure, i.e., $\varepsilon= 1$, has been solved in Ref. \cite{roszak15a}.
	That is, for any initially pure qubit state, the propagated state \eqref{sigma} is separable iff
	\begin{equation}
		\label{sep2}
		\hat{R}_{00}(t) = \hat{R}_{11}(t).
	\end{equation}
	Recall that the matrices $\hat{R}_{ii}(t)$ are density matrices that describe the state of the environment at time $t$ conditioned on the qubit pointer state $|i\rangle$, cf. Eq. \eqref{rij}.

	The above criterion is versatile since it does not require the knowledge of the full QE state at time $t$ for a given qubit-environment Hamiltonian.
	In fact, the QE Hamiltonian and the initial state of the environment contain all of the necessary information \cite{roszak15a}, and the actual conditional states of the environment do not have to be known.
	Note that, in general, distinguishing if a given state is entangled is a problem of the same complexity as finding the actual amount of entanglement of the state.

	The separability criterion in Eq.~\eqref{sep2} involves actual environmental states that can be observed.
	Specifically, schemes for the detection of QEE based on the criterion and, importantly, using operations and measurements performed only on the qubit have been proposed in Ref.~\cite{roszak19a,rzepkowski20i} and experimentally measured in Ref.~\cite{bizzarri25}.
	Moreover, the criterion has also been used to define a well-suited qubit-environment entanglement measure for pure dephasing evolutions that works for pure initial qubit states \cite{roszak20}.
	It allows for the quantification of this type of entanglement with the use of significantly less computing power \cite{strzalka24}.

\section{QEE for mixed initial qubit states}
\label{sec4}

\subsection{Separable evolutions}
\label{sec4a}

	If, at a given time $t>0$, the pure-state separability criterion \eqref{sep2} is satisfied, then it is straightforward to show that it holds true for an initially mixed qubit state \eqref{iniq}. 
	To show this, we substitute the explicit form of $R_{11}(t)$ into criterion \eqref{sep2},
	\begin{equation}
		\hat{R}_{00}(t) = \hat{w}(t)\hat{R}_{00}(t)\hat{w}^{\dagger}(t).
	\end{equation}
	Multiplying this from the right by $\hat{w}(t)$ yields
	\begin{equation}
		\label{sep3}
		\left[\hat{R}_{00}(t),\hat{w}(t)\right]=0
	\end{equation}
	since the operator $\hat{w}(t)$ is unitary as a product of two unitaries, cf. Eq. \eqref{w2}.
	A vanishing commutation relation means that there exists a basis in which both the conditional environmental density matrix $\hat{R}_{00}(t)$ and the environmental operator $\hat{w}(t)$ are diagonal.
	We denote this basis as $\{|n(t)\rangle\}$;
	thus, we can write
	\begin{align}
		\label{r00}
		\hat{R}_{00}(t)
		={}&
		\sum_n p_n(t)|n(t)\rangle\langle n(t)|,
		\\
		\label{wt}
		\hat{w}(t)
		={}&
		\sum_n e^{i\phi_n(t)}|n(t)\rangle\langle n(t)|.
	\end{align}
	In this basis, the qubit-environment density matrix at time $t$ in Eq. \eqref{sigma2} takes the form
	\begin{align}
		\label{sigma3}
		{}&\hat{\sigma}(t)
		\\\nonumber
		={}&\sum_n p_n(t)\left(
		\begin{array}{cc}
			|a|^2&ab^*\varepsilon e^{-i\phi_n(t)}
			\\
			a^*b\varepsilon e^{i\phi_n(t)}&|b|^2
		\end{array}
		\right)\otimes |n(t)\rangle\langle n(t)|,
	\end{align}
	which is clearly separable.
	Note that, similarly to the case of a pure initial qubit state, this does not mean the absence of qubit decoherence because tracing out the environment yields
	\begin{align}
		\label{rhoq}
		{}&\hat{\rho}(t)=\Tr_E\hat{\sigma}(t)
		\\
		\nonumber
		={}&
		\left(
		\begin{array}{cc}
			|a|^2&ab^*\varepsilon \sum_n p_n(t)e^{-i\phi_n(t)}
			\\
			a^*b\varepsilon \sum_n p_n(t)e^{i\phi_n(t)}&|b|^2
		\end{array}
		\right),
	\end{align}
	in which the diagonal elements of the density matrix remain constant while the amplitude of the off-diagonal elements is proportional to $|\sum_n p_n(t)e^{i\phi_n(t)}| \leq 1$.
	Furthermore, the separable state \eqref{sigma3} is also zero-discordant \cite{ollivier01,henderson01,modi14} with respect to the environment since the states $|n(t)\rangle$ form an orthonormal basis.
	This is also the case for pure initial qubit states;
	see Ref. \cite{roszak17i} for details.

\subsection{Entangling evolutions}
\label{sec4b}

	As in the original papers studying QEE \cite{roszak15a} and system-environment entanglement \cite{roszak18}, we here too start with the seminal positive partial transpose (PPT) criterion \cite{horodecki96i,peres96a_i}.
	It states that if any of the eigenvalues of a bipartite density matrix become negative after a partial transposition with respect to either subsystem, the state is entangled.
	Otherwise, it is separable, or so-called bound entanglement is present in the system \cite{horodecki97b,horodecki98}.
	However, for pure initial states of the qubit ($\varepsilon=1$), bound entangled states cannot be formed during pure dephasing evolutions \cite{roszak15a,horodecki00}.
	In the case of mixed initial qubit states ($\varepsilon<1$), such entanglement is generally possible, but nevertheless the present study focuses on mainly resourceful free (not bound) entanglement.

	For our analysis, we work with the qubit-environment density matrix in the form of Eq. \eqref{sigma2} and use the eigenbasis of the conditional density matrix of the environment $\hat{R}_{00}(t)$, $\{|n(t)\rangle\}$ as defined in Eq. \eqref{r00}.
	Note that the evolution operator $\hat{w}(t)$ now is not diagonal in this basis since the separability criterion, Eq. \eqref{sep3}, does not hold.

	First, we derive a set of QE entanglement criteria following path in Ref. \cite{roszak15a}.
	To this end, we start by performing the partial transposition of state \eqref{sigma2} with respect to the qubit;
	this is particularly straightforward with the notation being used since it only involves exchanging the off-diagonal blocks.
	The matrix, written explicitly in the eigenbasis of $\hat{R}_{00}(t)$, reads
	\begin{align}
		\label{sigmaT}
		&{}\hat{\sigma}^{T_Q}(t)
		\\
		\nonumber
		=&{}\left(
		\begin{array}{cc}
			|a|^2\sum\limits_n p_n|n\rangle\langle n|
			&
			a^*b\varepsilon\sum\limits_{k,n} p_ny_{kn}|k\rangle\langle n|
			\\
			ab^*\varepsilon\sum\limits_{k,n} p_ny_{kn}^*|n\rangle\langle k|
			&
			|b|^2\sum\limits_{k,q,n} p_ny_{kn}y_{qn}^*|k\rangle\langle q|
		\end{array}
		\right).
	\end{align}
	Here, we omitted the explicit time-dependence of $p_n$, $y_{kn}$, and $|n\rangle$ for clarity, which is also done in the continuation of this work.
	The coefficients $y_{kn}$ are given by
	\begin{equation}
		\label{y}
		y_{kn}(t)=\langle k(t)|\hat{w}(t)|n(t)\rangle.
	\end{equation}

\begin{widetext}
	In order to check if the matrix \eqref{sigmaT} has negative eigenvalues, it suffices to check if any of its principal minors are negative according to Sylvester's criterion.
	The study in Ref. \cite{roszak15a} restricted to a single class of minors, obtained by crossing $N-1$ rows and columns (here $N$ is the dimension of the environment) in such a way that only a single diagonal element proportional to $|b|^2$ is left.
	This resulted in
	\begin{equation}
		\label{Mi}
		M_i
		=
		\det\left(
		\begin{array}{cccc}
			|a|^2p_0&\dots&0&a^*b\varepsilon p_i y_{0i}
			\\
			\vdots&\ddots&\vdots&\vdots
			\\
			0&\dots&|a|^2p_{N-1}&a^*b\varepsilon p_i y_{N-1i}
			\\
			ab^*\varepsilon p_iy_{0i}^*&\dots&ab^*\varepsilon p_iy_{N-1i}^*&|b|^2\sum\limits_n p_n|y_{in}|^2
		\end{array}
		\right)
		=
		|a|^{2N}|b|^2
		\prod_k p_k
		\sum_{n: p_n\neq 0}\left(
			p_n|y_{in}|^2
			-\varepsilon^2\frac{p_i^2}{p_n}|y_{ni}|^2
		\right).
	\end{equation}
	For pure states, $\varepsilon=1$, all minors $M_i$ can only be negative or equal to zero;
	see Ref. \cite{roszak15a} for details.
	And the separability criterion in Sec. \ref{sec3} can be derived by a careful analysis of when the minors are not negative, in combination with a direct proof of separability in this case.
\end{widetext}

	For mixed initial qubit states ($0<\varepsilon<1$), the situation is much more complicated, however.
	We exclude the trivial case $\varepsilon=0$, when entanglement cannot be formed since the density matrix \eqref{sigma2} clearly admits a separable decomposition.
	The minors \eqref{Mi} can now be positive such that there are situations when, in principle, entangling evolutions are not accompanied by entanglement generation.
	Furthermore, in situations when all minors are non-negative because of the initial mixedness of the qubit, it is not possible to explicitly show separability of the QE density matrix.
	In fact, such a density matrix does not have to be separable because there are other principal minors outside of the set $M_i$ that could be negative.
	Thus, the family of minors \eqref{Mi} now only witnesses entanglement \cite{terhal00,chruscinski14}.
	That is, if any of the minors is negative, then there is entanglement in the QE density matrix\JS;
	but if all of them are non-negative, then no information about entanglement or separability is obtained.
	Thus, two important questions are not resolved at this stage through prior works:
	Is sudden death of entanglement possible for mixed initial qubit states?
	Can QEE exist in entangling evolutions if all minors $M_i$ are non-negative?
	Neither is possible for pure qubit states.

	In order to explore possible entanglement generation when it is in principle possible, but it is not signified by the witnesses $M_i$, we employ positive transformations on the partially transposed matrix \eqref{sigmaT} to reduce the dimension of the diagonalization problem that needs to be solved to check the PPT criterion.
	For this purpose, we note that a matrix of the form
	$\left(
	\begin{smallmatrix}
	\hat{A}&\hat{C}\\
	\hat{C}^{\dagger}&\hat{B}
	\end{smallmatrix}
	\right)$, where $\hat{A}$, $\hat{B}$, and $\hat{C}$ are square matrices themselves, can be transformed as follows without changing the signs of its eigenvalues:
	\begin{equation}
		\hat{U}_2\hat{U}_1\left(
		\begin{array}{cc}
			\hat{A}&\hat{C}
			\\
			\hat{C}^{\dagger}&\hat{B}
		\end{array}
		\right)\hat{U}_1^{\dagger}\hat{U}_2^{\dagger}
		=
		\left(
		\begin{array}{cc}
			\mathbbm{1}&0
			\\
			0&\hat{D}
		\end{array}
		\right),
	\end{equation}
	with
	$
		\hat{D}=\hat{B}-\hat{C}^{\dagger} \hat{A}^{-1} \hat{C}
	$, $
		\hat{U}_1= \left(
		\begin{smallmatrix}
		\hat{A}^{-1/2}&0\\
		0&\mathbbm{1}
		\end{smallmatrix}
		\right)
	$, and $
		\hat{U}_2= \left(
		\begin{smallmatrix}
			\mathbbm{1}&0\\
			-\hat{C}^{\dagger}\hat{A}^{-1/2}&\mathbbm{1}
		\end{smallmatrix}
		\right)
	$.
	Thus, one can probe the positivity of matrix $\hat{D}$ instead of the full matrix, reducing the dimension of the diagonalization by a factor of two.
	Even if this is not a strong reduction of the numerical diagonalization problem when dealing with large environments, there is an analytical relationship between the matrix $\hat{D}$ and the family of minors $M_i$ for the problem at hand, which we demonstrate presently.

	The matrix $\hat{D}$ can be written explicitly using the partially transposed matrix \eqref{sigmaT},
	\begin{equation}
		\label{D1}
	\begin{aligned}
		\hat{D}
		={}&
		|b|^2\left(
			\hat{w}(t)\hat{R}_{00}(t)\hat{w}^{\dagger}(t)
		\right.
		\\
		{}&\left.
			-\varepsilon^2\hat{R}_{00}(t)\hat{w}^{\dagger}(t)\hat{R}_{00}^{-1}(t)\hat{w}(t)\hat{R}_{00}(t)
		\right).
	\end{aligned}
	\end{equation}
	Applying the spectral decomposition of $\hat{R}_{00}(t)$ further yields
	\begin{equation}
		\label{D2}
	\begin{aligned}
		\hat{D}
		={}&
		|b|^2\sum_{k,q}\left(
			\sum_np_ny_{kn}y_{qn}^*
		\right.
		\\
		{}&\left.
			-\varepsilon^2p_kp_q\sum_{n:p_n\neq 0}p_n^{-1}y_{nq}y_{nk}^*
		\right)|k\rangle\langle q|.
	\end{aligned}
	\end{equation}
	For entangling evolutions, this matrix is not diagonal, and its eigenvalues can---for the most part, and except for an environment composed of a single qubit---only be found numerically.
	Nevertheless, a meaningful observation can be made here:
	the diagonal elements of matrix \eqref{D2} for $k,q=i$ are proportional to the corresponding minor $M_i$;
	in fact, they are the same functions of time up to a constant that only depends on the initial states of the qubit and the environment.
	This means that the eigenvalues of the matrix $\hat{D}$ do not correspond to the minors $M_i$, and thus can cross zero at different points in time. Obviously, when any minor $M_i$ is negative, the matrix $\hat{D}$ must have a negative eigenvalue, but the PPT criterion can be broken even when the set of specific minors $M_i$ does not indicate entanglement.

\begin{figure}
	\includegraphics[width=1\linewidth]{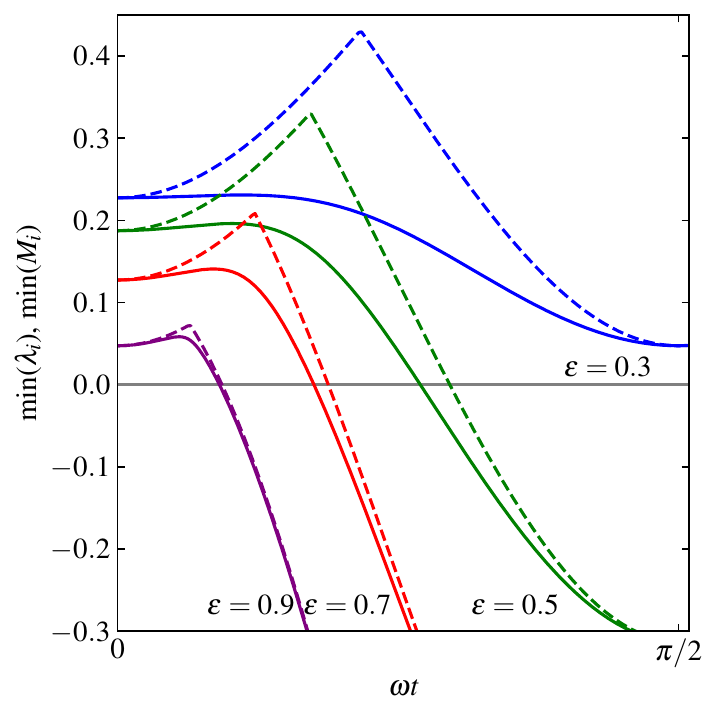}
	\caption{%
		Evolution of normalized entanglement witnesses for a proof-of-concept single-qubit environment for different initial coherence parameters, $\varepsilon\in\{ 0.3,0.5,0.7,0.9\}$.
		Solid lines show $\min_i\lambda_i=\lambda_-$, and dashed lines depict $\min_i M_i$.
		Entanglement is present whenever the solid lines are below zero.
	}\label{kubit}
\end{figure}

	To show that the aforementioned scenario does in fact happen, we now turn to the simplest situation when the environment consists of only a single qubit.
	In this case, the eigenvalues of the two-dimensional matrix $\hat{D}$ can be found analytically and are given by $
		\lambda_{\pm}=\frac{1}{2}\left(
		D_{00}+D_{11}\pm\sqrt{\Delta}
		\right)
	$, with $\Delta=(D_{00}-D_{11})^2+4|D_{01}|^2$ defined through the matrix entries $D_{ij}$.
	The smaller eigenvalue $\lambda_-$ can be compare to the minimum of the two minors $M_i$ in Eq. \eqref{Mi}.
	Figure \ref{kubit} shows the evolution of the minimum eigenvalue of $\hat{D}$ (solid lines) and $\min{M_i}$ (dashed lines) for the evolution operator \eqref{w2}, here
	\begin{equation}
		\hat{w}(t)=e^{i\omega t}|+\rangle\langle +|+e^{-i\omega t}|-\rangle\langle -|,
	\end{equation}
	with $|\pm\rangle = (|0\rangle\pm |1\rangle)/\sqrt{2}$ and initial state of the environment $\hat{R}(0)=p_0|0\rangle\langle 0|+p_1|1\rangle\langle 1|$ for $p_0=3/4$.
	The curves are normalized such that both types of curves coincide at $t=0$.
	Then, $\lambda_-<0$ is an if-and-only-if condition for free entanglement, while $\min_{i} M_i <0$ is only an if condition.

	In Fig. \ref{kubit}, we immediately observe a number of behaviors of the evolution that are impossible for pure initial qubit states.
	First, during an evolution that is entangling in its nature, the QE density matrix can remain separable throughout, as seen for $\varepsilon =0.3$.
	The requirement here is that the initial state of the qubit is strongly mixed, i.e., close to the maximally mixed state.
	Second, sudden birth and sudden death of entanglement is observed;
	note that the behavior is symmetric about the $\omega t=\pi/2$ axis.
	Such effects are not possible for pure initial qubit states, and only isolated points of separability can occur during an entangling evolution \cite{roszak17i}.
	The third observation is that the two types of functions are distinct, and entanglement verified through $\lambda_-<0$ is present for longer time periods than observed by the minors.
	Nevertheless, for larger environments, the study of the class of minors $M_i$ is more practical than finding eigenvalues of a matrix of the same dimension as the environment.
	Still, it is important to keep in mind that minors might fail for small amounts of entanglement.

\section{Application to an NV center spin qubit in an environment of nuclear spins}
\label{sec5}

\begin{figure*}
	\includegraphics[width=1.\linewidth]{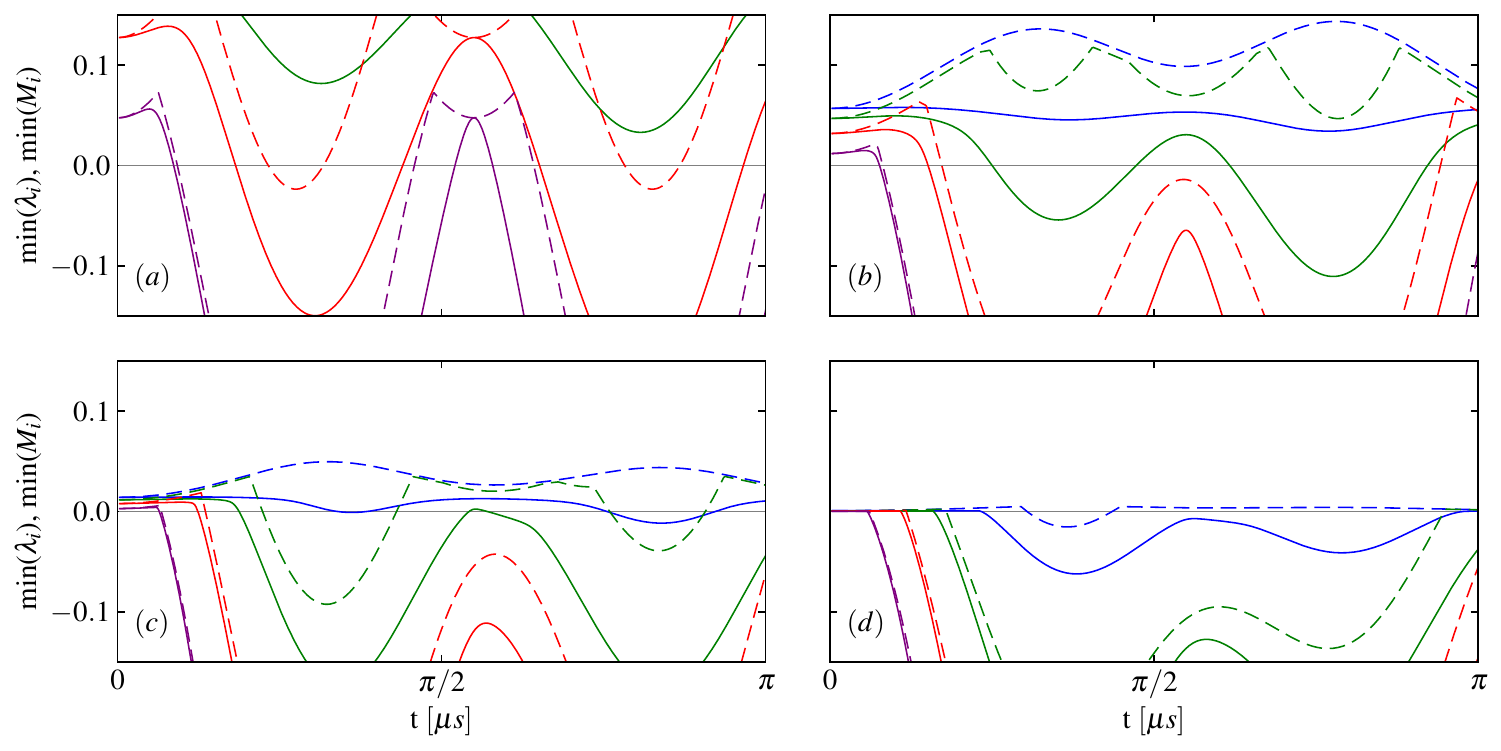}
    \caption{%
		Evolution of normalized entanglement witnesses for an NV center spin qubit interacting with (a) one, (b) two, (c) three, and (d) five environmental spins for different initial qubit coherence, $\varepsilon\in\{0.3,0.5,0.7,0.9\}$.
		Solid lines depict $\min_i \lambda_i$, and dashed lines show $\min_i M_i$.
		The initial polarization of each environmental spin is $p=0.5$, and the applied magnetic field is $B_z=0.1\,\mathrm{T}$.
		Colors correspond to those in Fig. \ref{kubit}.
	}\label{rysNV}
\end{figure*}

	To provide an example of how the two types of criteria---minors and minimal eigenvalues---behave in realistic environments, we now study an NV center spin qubit interacting with an environment of nuclear spins via the hyperfine interaction \cite{jelezko06, gali08, zhao12_d, doherty13, kwiatkowski18, wood18i, strzalka20, strzalka21,finsterhoelzl25}.
	The NV center spin qubit is a defect in the diamond crystal lattice in which two neighboring carbon atoms are replaced by a nitrogen atom and a vacancy, respectively.
	The lowest energy states of an NV center have spin $S = 1$, and they form a qutrit with uneven energy spacing;
	thus, one can choose any two states to use as a spin qubit \cite{Wrachtrup_JPC06}.
	In the following, we are going to use a qubit defined on the $S_\mathrm{z}=0,1$.
	The environments in diamond tend to be sparse \cite{jelezko06, doherty13}, but we specifically limit the study to a few nuclear spins in order to witness interesting behaviors of entanglement.

	The Hamiltonian of the full $S = 1$ qutrit consists of three parts.
	The free Hamiltonian of the qutrit has no bearing on QEE, so we do not take it into account.
	The free Hamiltonian of the environment is given by
	\begin{align}
		\hat{H}_E=\sum_{k}\gamma_{n}B_{z}\hat{I}^{z}_{k}.\label{hamE}
	\end{align}
	with the gyromagnetic ratio set to $\gamma_{n} = 10.71\,\mathrm{MHz/T}$, the magnetic field $B_z$ applied in the $z$ direction, and operators of the $z$ component of nuclear spin $k$, denoted by $\hat{I}^{z}_{k}$.
	The interaction term is of the form
	\begin{equation}
		\label{HQE}
		\hat{H}_{QE}= -\ket{-1}\bra{-1} \otimes\hat{V} +\ket{1}\bra{1} \otimes\hat{V},
	\end{equation}
	where the operator $\hat{V}$ is given by
	\begin{equation}
		\label{V}
		\hat{V} = \sum_{k}\sum_{j \in (x,y,z)} \mathbb{A}^{z,j}_k\hat{I}^{j}_{k}.
	\end{equation}
	Therein, the coupling constants are given by
	\begin{equation}
		\mathbb{A}^{z,j}_k=\frac{\mu_0}{4\pi}\frac{\gamma_e\gamma_n}{r_{k}^3}\left(1-\frac{3(\mathbf{r}_{k}\cdot\hat{\mathbf{j}})(\mathbf{r}_{k}\cdot\hat{\mathbf{z}})}{r_{k}^2}\right),
	\end{equation}
	where $\mu_0$ is the magnetic permeability of the vacuum, $\mathbf{r}_{k}$ denotes a displacement vector between the $k$\textsuperscript{th} nucleus and the qubit, and $\hat{\mathbf{j}}=\hat{\mathbf{x}}$, $\hat{\mathbf{y}}$ and $\hat{\mathbf{z}}$ are versors (unit vectors) in three orthogonal directions.
	For nonzero magnetic fields, the Hamiltonian of the environment \eqref{hamE} and the interaction term \eqref{HQE} do not commute;
	thus, the environmental free evolution is relevant for the generation of QEE and cannot be omitted in the conditional evolution operators of the environment, Eq. \eqref{w}.

	Since we do not take into account the interaction between the environmental spins, the evolution operators can be written in product form with respect to the different spins, $\hat{w}_{i}(t)=\bigotimes_{k}\hat{w}_{i}^k(t)$.
	We further assume that there are no initial correlations between the nuclear spins, so the initial environmental state too can also be written in product form, $\hat{R}(0)=\bigotimes_{k} \hat{\rho}_k$.
	This significantly simplifies the treatment of the qubit-environment density matrix.
	We suppose that each spin of the environment is initially polarized, so that the density matrix of $k$th environmental spin is given by
	\begin{equation}
		\hat{\rho}_k =\frac{1}{2}(1 + p_{k}\hat{I}^{z}_{k})\label{ro_k},
	\end{equation}
	with equal polarization for each spin, $p_k=0.5$.

	Figure \ref{rysNV} shows the evolutions of the entanglement witnesses for the NV-center spin qubit for different numbers of environmental spins.
	In addition, Table \ref{tab} contains the specific distances between the spin qubit and each individual environmental spin together with the corresponding coupling constants.
	Figure \ref{rysNV}(a) shows the evolutions while only first spin ($k=1$) is taken into account, Fig. \ref{rysNV}(b) includes the first two ($k=1,2$), Fig. \ref{rysNV}(c) the first three, and Fig. \ref{rysNV}(d) all five nuclear spins.
	As before, solid lines denote the minima of the eigenvalues of the matrix \eqref{D2}, and dashed lines plot the minima of the class of principal minors under study.
	The color-coding of the curves also corresponds to the different initial coherences of the qubit as in Fig. \ref{kubit}.

\begin{table}[b]
	\begin{tabular}{ccccc}
		\hline
		\hline
		$k$ & $r_k\,\mathrm{[nm]}$ & $\mathbb{A}^{z,x}_k\,\mathrm{[\frac{1}{\mu s}]}$ & $\mathbb{A}^{z,y}_k\,\mathrm{[\frac{1}{\mu s}]}$ & $\mathbb{A}^{z,z}_k\,\mathrm{[\frac{1}{\mu s}]}$
		\\
		\hline
		$1$&  $0.563961$&     $ 0.196941$	&$1.02333$	&$0.417774$\\
		\hline
		$2$&$  0.617788$&$ 0.499393$&$	0$&	$-0.353124$\\
		\hline
		$3$&$ 0.636801$&$	0.362104$&$	-0.627183$&$	0.284496 $\\
		\hline
		$4$&$  0.636801	$&$0.268226$&$-0.46458	$&$-0.246563 $\\
		\hline
		$5$&$0.667287	$&$0.339684$&$	-0.147088$&$	-0.300241$\\
		\hline\hline
	\end{tabular}
	\caption{%
		Distances of each nuclear environmental spin from the NV-center spin qubit and the corresponding coupling constants for the realizations  used to get Fig. \ref{rysNV}.
	}\label{tab}
\end{table}

	Although the complexity of the evolution of the different curves increases with the number of environmental spins, the important feature is when the curves are positive or negative, which is much simpler for five environmental spins than it is for one.
	The evolution for a single qubit environment (a) is qualitatively similar to the example depicted in Fig.~\ref{kubit}, and the main difference is that QEE is much rarer---the NV center does not entangle at all also for $\varepsilon=0.5$---because the interaction is not capable of generating very strong entanglement.
	Yet, for a larger number of spins, we have situations where the prediction of the entanglement witness based on minors is qualitatively wrong.
	This is already visible for two qubits (b) for $\varepsilon = 0.5$, for which the minors suggest that the evolution is separable, while a repeated birth and death of entanglement occurs.
	This is also true for the $\varepsilon =0.3$ curve for a three-spin environment (c).
	The discrepancy of the $\varepsilon =0.3$ curve for the five-spin environment (d) is different.
	After the initial sudden birth of entanglement occurs, entanglement is continuously present in the system, but the minors indicate entanglement only in a singular, short time window.

	The application presented in this section shows that although the minimum of the class of minors is an entanglement witness, it is not suitable for the study of the behavior of QEE, not even qualitatively.
	This is particularly true when the amount of generated entanglement is small;
	recall in this context that the eigenvalues of the matrix \eqref{D2} hold the same information as the entanglement measure negativity \cite{Plenio_QIC07}, so they are enough to judge the amount of entanglement.

\section{Conclusion}
\label{sec6}

	We have studied the applicability to mixed states of the conditions that constitute a set of if and only if conditions of separability between a qubit and its environment as long as the joint evolution can only lead to pure dephasing of the qubit and the qubit is initially in a pure state.
	For mixed initial states of the qubit, these conditions only form an entanglement witness;
	i.e., if any of the conditions is broken, there is entanglement present, but the state can be entangled also when none of the conditions are violated.
	We showed this using the PPT criterion with the partially transposed qubit-environment density matrix transformed in such a manner that finding its negative eigenvalues only required the diagonalization of a matrix of the dimension of the environment.

	Furthermore, we have shown that a number of behaviors in the evolution of entanglement not allowed when the initial qubit state is pure are rendered possible for the initially mixed case.
	Most notably, this includes sudden birth and sudden death of entanglement.
	An initial period of separable joint evolution before entanglement occurs universally for all of the scenarios studied.
	This is predicted unambiguously by the conditions based on minors, but translates to the actual behavior of entanglement in all of the studied examples.

	Notably, evolutions that are entangling for pure qubit states can lead to no generation of entanglement throughout the evolution.
	This is observed for strongly decohered qubits or for interactions that are capable only of generating small amounts of entanglement.
	In such situations, proper measures of entanglement have to be used since witnesses can easily overlook its presence.
	Along the same line, partial transposition may be replaced with another positive but not completely positive map to detect additional entanglement.


\begin{acknowledgments}
	This project was funded within the QuantERA II Programme that has received funding from the EU H2020 research and innovation program,
	under Grant Agreement No 101017733, and with funding organisation MEYS (The Ministry of Education, Youth and Sports) of the Czech Republic and DPG (German Physical Society). MS acknowledges support by the Grant Agency of Charles University (122124).
\end{acknowledgments}

\end{document}